\documentstyle[11pt,emulateapj]{article}

\slugcomment{}

\newcommand{\etal}{{\it et\thinspace al.}\ }
\newcommand{\kms}{km\thinspace s$^{-1}$}
\newcommand{\simlt}{\ {\raise-.5ex\hbox{$\buildrel<\over\sim$}}\ }

\lefthead{Salzer \etal}
\righthead{KPNO International Spectroscopic Survey}

\begin{document}
\submitted{Accepted to the A.J.}
\title{The KPNO International Spectroscopic Survey.  \\I. Description of the Survey}

\author{John J. Salzer\altaffilmark{1,2} and Caryl Gronwall\altaffilmark{1,3}}
\affil{Astronomy Department, Wesleyan University, Middletown, CT 06459; 
slaz@astro.wesleyan.edu, caryl@adcam.pha.jhu.edu}

\author{Valentin A. Lipovetsky\altaffilmark{1,4} and Alexei Kniazev\altaffilmark{1}}
\affil{Special Astrophysical Observatory, Russian Academy of Sciences, Nizhny Arkhyz, Karachai-Circessia 357147, Russia; akn@sao.ru}

\author{J. Ward Moody}
\affil{Department of Physics \& Astronomy, Brigham Young University, Provo, UT 84602; jmoody@astro.byu.edu}

\author{Todd A. Boroson}
\affil{U.S. Gemini Program, National Optical Astronomy Obs., P.O. Box 26732, Tucson, AZ 85726; tyb@noao.edu}

\author{Trinh X. Thuan}
\affil{Astronomy Department, University of Virginia, Charlottesville, VA 22903; txt@starburst.astro.virginia.edu}

\author{Yuri I. Izotov}
\affil{Main Astronomical Observatory, National Academy of Sciences of Ukraine, Goloseevo, Kiev 03680, Ukraine; izotov@mao.kiev.ua}

\author{Jose L. Herrero}
\affil{BBN Technologies, Cambridge, MA 02140; jose@world.std.com}

\author{Lisa M. Frattare\altaffilmark{1}}
\affil{Space Telescope Science Institute, Baltimore, MD 21218; frattare@stsci.edu}

\altaffiltext{1}{Visiting Astronomer, Kitt Peak National Observatory. 
KPNO is operated by AURA, Inc.\ under contract to the National Science
Foundation.} 
\altaffiltext{2}{NSF Presidential Faculty Fellow.} 
\altaffiltext{3}{present address: Department of Physics \& Astronomy, Johns Hopkins University,
Baltimore, MD 21218.}
\altaffiltext{4}{Deceased 22 September 1996.} 

\begin{abstract}
The KPNO International Spectroscopic Survey (KISS) is a new objective-prism
survey for extragalactic emission-line objects.  It combines many of the features
of previous slitless spectroscopic surveys that were carried out with Schmidt telescopes
using photographic plates with the advantages of modern CCD detectors.  It is the
first purely digital objective-prism survey, and extends previous photographic
surveys to substantially fainter flux limits.  In this, the first paper in the
series, we give an overview of the survey technique, describe our data processing 
procedures, and present examples of the types of objects found by KISS.  Our first
H$\alpha$-selected survey list detects objects at the rate of 18.1 per square
degree, which is 181 times higher than the surface density of the Markarian survey.
Since the sample is line-selected, there is an imposed redshift limit of z$\simlt$0.095
due to the filter employed for the objective-prism observations.  We evaluate the
quality of the observed parameters derived from the survey data, which include
accurate astrometry, photometry, redshifts, and line fluxes.  Finally, we describe
some of the many applications the KISS database will have for addressing specific
questions in extragalactic astronomy.  Subsequent papers in this series will 
present our survey lists of emission-line galaxy candidates.
\end{abstract}



\clearpage
\section{Introduction}

\par Active galactic nuclei (AGN) and galaxian starbursts are among the
most energetic phenomena known in the universe.  From Blue Compact Dwarf 
galaxies to QSOs, galaxian activity manifests itself on all scales and 
over the entire range of the electromagnetic spectrum.  Activity in galaxies
appears to be common, with between 5 and 10\% of all galaxies showing
evidence for it via the presence of unusually strong emission lines in
their spectra (e.g., Salzer 1989, Gregory \etal 2000).  The study of these 
objects has been a major research
topic for many years, and a great deal has been learned about the physical
processes occurring in the active regions of both AGN and starburst
galaxies.  Much still remains to be understood.  How does the activity,
be it central or global, shape the overall evolution of the host galaxy?
Do all galaxies cycle through periods of activity?  What role does 
environment play in whether or not a galaxy becomes active?  Given that
many of the galaxies we observe at high redshift are active, can we 
understand their evolutionary status in the context of what we observe
locally?  How has the global star formation density evolved as a function of
cosmic epoch?

\par Central to the study of AGN and starburst activity have been the surveys
which have cataloged large samples for subsequent investigation.  Few types of 
extragalactic surveys have been as scientifically fruitful as the various 
objective-prism surveys for UV-excess and emission-line galaxies (ELGs) carried 
out with Schmidt telescopes.  Much of what we know about Seyfert galaxies, 
starburst galaxies, and even QSOs has been learned by studying objects originally 
discovered in surveys with such familiar names as the Markarian, Tololo, 
Wasilewski, Michigan (UM), Kiso, Case, and Second Byurakan (SBS) surveys.

\par Virtually all of the existing surveys for galaxies which display some
form of unusual activity (e.g., blue colors or strong emission lines) have
been carried out using Schmidt telescopes and one of three detection
methods.  The first, which is commonly referred to as the color-selection
technique, uses multiple exposures taken through two or three filters
to isolate the bluest galaxies.  This method has been used by Haro (1956), 
Takase \& Miyauchi-Isobe (1993; Kiso survey), and Coziol {\it et al.} (1993, 
1997; Montreal survey).  The well known Palomar-Green survey (Green {\it et al.} 
1986; PG survey) for UV-bright stars and QSOs also employed this technique.  
The other two methods employ objective prisms.  One, the 
UV-excess technique, was pioneered by Markarian (1967) and Markarian \etal (1981).  
It is similar to the color-selection method in that the criterion for inclusion is 
the presence of a very blue continuum.  The other selects objects via the presence 
of emission lines in the objective-prism spectra.  Surveys of this type include 
the Tololo (Smith 1975, Smith {\it et al.} 1976), Michigan (UM, MacAlpine {\it et al.} 
1977, MacAlpine \& Williams 1981), Wasilewski (1983), and Universidad Complutense de 
Madrid (UCM, Zamorano {\it et al.} 1994, 1996; Alonso {\it et al.} 1999) surveys.  
Finally, the Case (Pesch \& Sanduleak 1983, Stephenson {\it et al.} 1992) and Second 
Byurakan surveys (Markarian \etal 1983, Markarian \& Stepanian 1983, Stepanian 1994) 
use a hybrid scheme, selecting galaxies based on both UV excess and line emission.

\par In recent years various observatories have begun experimenting with 
the use of CCD detectors on Schmidt telescopes.  While CCDs had become 
the detector of choice for most observing applications by the mid-1980's,
photographic plates have remained in use on Schmidt telescopes due to their 
vastly superior areal coverage.  However, the advent of large format 
CCDs (2048 $\times$ 2048 pixels and larger) now provides sufficiently 
large fields-of-view (in excess of one degree square) that it has become 
advantageous to carry out a variety of wide-field survey projects using CCDs 
on Schmidt telescopes (Armandroff 1995).  These advances have motivated us 
to initiate a new survey for emission-line galaxies. 

\par Our new survey is called KISS - the KPNO International Spectroscopic 
Survey.  The technique we employ combines the benefits of a traditional 
photographic objective-prism survey with the advantages of using a CCD 
detector, and represents the {\bf next generation of wide-field 
objective-prism surveys}.  The goal of the KISS project is to survey a 
large area of the sky for extragalactic emission-line sources, and 
to reach a minimum of two magnitudes deeper than any of the previous 
line-selected Schmidt surveys.  Our initial plan was to be able to detect 
strong-lined ELGs with continuum magnitudes down to B = 20 - 21, and to 
be complete to B = 19 - 20.  The use of a CCD detector allows us to achieve 
these goals.  

\par We stress that the need for this type of survey goes beyond the desire 
to simply find {\bf more} AGN and starburst galaxies.  The type of scientific
questions we wish to address requires large samples with well defined selection 
criteria and completeness limits.  Our research cuts across the traditional
lines of study of active galaxies, and touches upon nearly all areas of
extragalactic astronomy, such as galaxy formation and evolution, chemical 
evolution, and large-scale structure and cosmology.  Although we fully expect 
to discover many new objects which possess noteworthy characteristics that will 
warrant individual study, the prime motivation for this survey is to create
a sample of objects that can be used as probes for studying more general
questions.  We will discuss some of these applications of the KISS sample
of galaxies in Section 5.

\par In the next section we describe in detail our instrumental setup and 
observational technique, while Section 3 gives an overview of our image 
processing and analysis software which is central to the survey effort.  
Section 4 presents the results from our first two survey lists,
evaluates the quality of the survey data (e.g., photometry, astrometry,
redshifts), and gives a preliminary analysis of the survey contents.
In Section 5 we list some of the possible uses of the KISS sample
for attacking specific problems relevant to extragalactic astronomy and 
cosmology.  Finally, Section 6 summarizes our results.


\section{Observations}

\subsection{Instrumentation}

All survey data were acquired using the 0.61-meter Burrell Schmidt 
telescope\footnote{Observations made with the Burrell Schmidt of the
Warner and Swasey Observatory, Case Western Reserve University.  Prior to
1997 October, the Burrell Schmidt was operated jointly by CWRU and KPNO.} 
located on Kitt Peak. 
During the observations of the first survey strip, the detector used
was a 2048 $\times$ 2048 pixel STIS CCD (S2KA).  This CCD has 21-$\micron$
pixels, which yielded an image scale of 2.03 arcsec/pixel.  The overall
field-of-view was 69 $\times$ 69 arcmin, and each image covered 1.32 
square degrees.  The CCD was illuminated fairly evenly by the telescope,
although the slightly undersized Newtonian secondary mirror introduced
modest vignetting (10--15\%) on the edges of the images.

The spectral data were taken with two separate prisms installed on the
top ring of the telescope, in front of the corrector plate.  Data taken
in the blue portion of the spectrum utilized a 2$\arcdeg$ prism, which
gave spectra with a reciprocal dispersion of 19 \AA/pixel at 5000 \AA.
The red survey spectral data were obtained with a 4$\arcdeg$ prism,
which provided a reciprocal dispersion of 24 \AA/pixel at H$\alpha$.
Thus, the two sets of spectral data have roughly the same spectral
resolution.

During the planning stages for the survey, it was deemed essential to 
restrict the wavelength range covered by the objective-prism spectra.
Although traditional objective-prism surveys typically did not employ
filters, there are two primary reasons why it is important for our
CCD-based survey.  First, CCDs are more-nearly panchromatic detectors
than are photographic plates.  For example, the IIIa-J emulsion used
for many of the previous photographic surveys has a sensitivity cutoff 
above 5350 \AA,
effectively limiting the length of each spectrum on the red side.  No
such cutoff exists for CCDs.  Second, and more important for the
present survey, is the depth to which CCDs can reach relative to 
photographic plates.  Since our CCD can detect continua in objects as
faint as B = 20 -- 21, the problem of image crowding and overlap
becomes enormous.  An unfiltered image taken with the 2$\arcdeg$ prism
yields spectra that are $\sim$250 pixels long!  With typically 4000 -- 
10,000 objects per image, essentially every spectrum is overlapped 
by another.  For the photographic surveys, which rarely detected the
continuum in objects fainter than B = 17 or 18, spectral overlap was
not as severe a problem.

Therefore, two specially designed filters were used for the KISS
observations in order to restrict the wavelength range of the spectra.
The filter for the blue spectral survey was designed to cover the
wavelength range from just below restframe H$\beta$ up to, but not 
including, the strong [O I]$\lambda$5577 night sky line.  The decision
to exclude the night sky [O I] line was based in large part on our
desire to keep the sky background as low as possible.  Since in a
slitless spectroscopic survey like KISS, every pixel of the detector
is illuminated by the night sky spectrum {\it at all wavelengths},
the sky background can pose a serious limitation on the depth of the
survey data.  Hence, eliminating the bright [O I] line results in a
substantial reduction in the background flux.  The region covered
by our blue filter (see Figure 1a), 4800 -- 5520 \AA, is one of the 
darkest portions of the night sky spectrum.  The red spectral filter
(Figure 1b), was likewise chosen to have a blue limit just below
restframe H$\alpha$, and to extend up to the beginning of a strong
OH molecular band near 7200 \AA.  In both cases, the primary emission
line we are sensitive to ([\ion{O}{3}]$\lambda$5007 in the case of the
blue survey, and H$\alpha$ for the red survey) is detectable up to a
redshift of approximately 0.095.

\begin{figure*}[ht]
\plotone{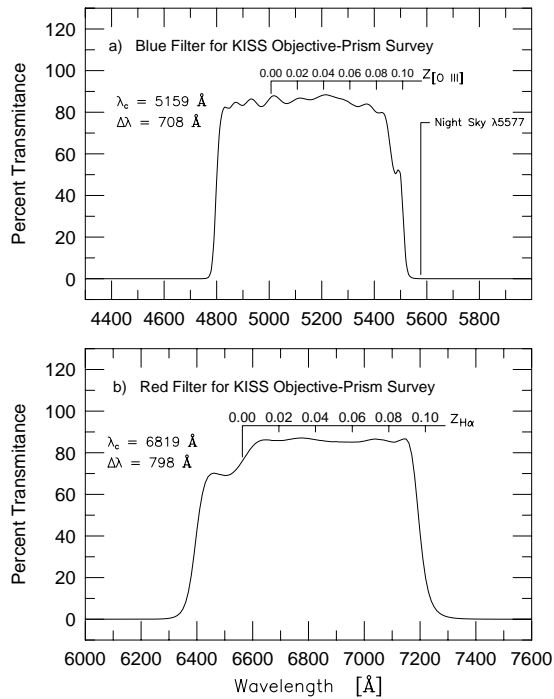}
\figcaption[fig1.eps]{Filter throughput tracings for the filters used in
the KISS project.  (a) The blue survey ([\ion{O}{3}] selected) filter.  (b) The 
red survey (H$\alpha$ selected) filter.  The central wavelengths ($\lambda_c$)
and widths at 50\%\ throughput ($\Delta\lambda$) are listed for both
filters.  The scale above the tracings shows the approximate location of
the relevant emission line for the redshifts indicated.  Both filters are
sensitive to their primary emission line out to redshifts of z $\approx$
0.095.  In addition, the blue filter is sensitive to H$\beta$ emission
out to z $\approx$ 0.13.\label{fig:fig1}}
\end{figure*}
\suppressfloats{}

\subsection{The Observational Data}

Three types of images are obtained for the survey.  The data for each 
survey field consists of direct images acquired through Harris B and V
filters (which closely mimic standard Johnson B and V filters), spectral 
images in one or both of the spectral regions mentioned above and taken 
through the relevant prism and spectral filter combination, and a pair 
of short exposure calibration images.  Examples of a direct and spectral 
image for a portion of one KISS field are shown in Figures 2 and 3, 
respectively.

{\it Direct Images}.  Roughly 30\% of the survey observing time is spent 
in direct imaging mode, i.e., with no prism on the telescope.  For each
field in the survey, we obtain direct images through two filters (B and V)
to a depth of 1 to 2 magnitudes fainter than the limiting 
magnitude of the spectral images.  This is done for several reasons:

\noindent (1) {\it Object location}.  It is difficult to assign accurate
positions to objects based on their objective-prism spectra.  The direct
images yield much more accurate positions, and can be used to locate 
objects that are substantially fainter than are visible in the spectral 
images.

\noindent (2) {\it Photometry}.  The direct images provide accurate B and 
V magnitudes for all objects in the fields. No follow-up imaging photometry 
is required.

\noindent (3) {\it Morphology}.  Likewise,
the direct images are crucial for assessing the nature of candidates in
the spectral images.  Although the resolution is poor, due to the coarse
pixel scale, the images are suitable for object classification (star vs.
galaxy) for at least the brighter objects (V $<$ 19).

\noindent (4) {\it Faint ELG detection}.
Without the direct images, only objects with a detectable continuum
could be selected from the objective-prism images by our automated
reduction procedure.  Faint ELGs with strong emission lines but weak continua
would not be detected, even if the line strengths were well above our
detection threshold.  By using object locations based on the deeper
direct images, even very faint ELGs (B $\approx$ 21) can be cataloged.

{\it Spectral Images}.  At the heart of the KISS project are the 
objective-prism spectral images.  The spectral data acquired for the 
survey are quite deep, between 1 and 2 magnitudes deeper than previous
line-selected samples like UM, Case, and UCM, and 3 to 4 magnitudes
deeper than the Markarian survey data.  As mentioned above, spectral
data have been obtained for different portions of the first survey
strip in one or both of two spectral regions.  The blue spectra were
obtained for all 102 survey fields, while the red spectra were taken
for 54 fields.  We began the survey using blue spectra exclusively, with
the idea of better matching many of the previous surveys.  However,
we carried out tests in the red spectral region during our second season of
observing, and realized that the combination of higher CCD sensitivity
in the red plus the more ubiquitous presence of strong H$\alpha$ (as opposed
to [\ion{O}{3}]) combined to more than compensate for the higher sky background
levels in the red.  The net result was a higher detection rate of ELGs 
per unit area in the red spectra.  Hence, after finishing the first
survey strip in the blue region, we reobserved as many fields as possible
in the red.  For all subsequent survey strips, we plan to observe in the red 
spectral region only, as the red survey detects nearly all (at least 87\%)
of the objects seen in the blue survey (see Section 4).

\begin{figure*}[ht]
\plotone{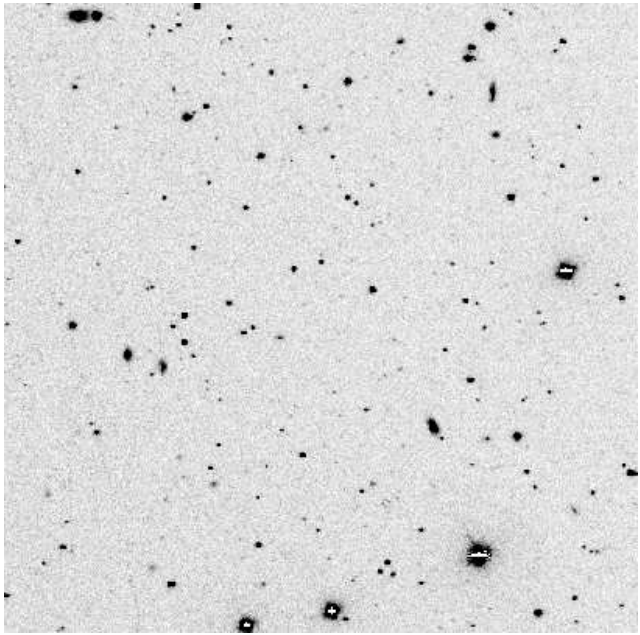}
\figcaption[fig2.eps]{An example direct image.  This is a small section 
of the combined (B + V) direct image of survey field F1305.  The area shown 
represents a 512 $\times$ 512 pixel subsection of the total field, or 6.25\% 
of the area of F1305.\label{fig:fig2}}
\end{figure*}

\begin{figure*}[ht]
\plotone{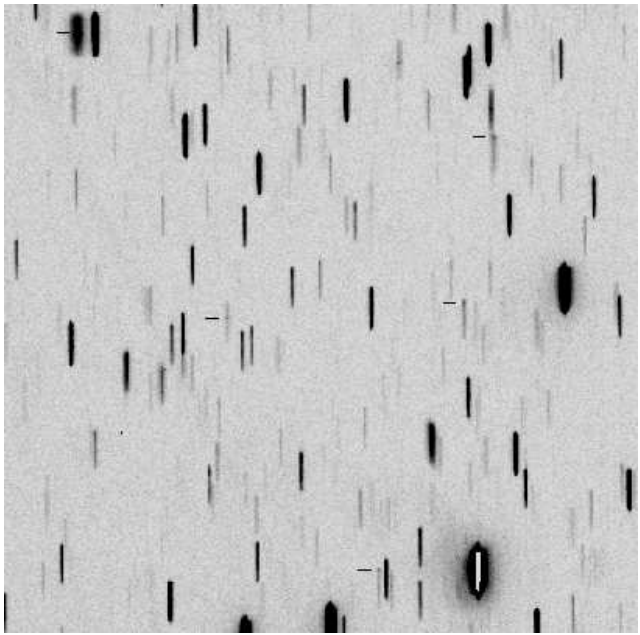}
\figcaption[fig3.eps]{An example spectral image.  The area shown is the
same region depicted in Figure 2, but for the red spectral data.  
Eight emission-line galaxy candidates located in this field are indicated by 
the tick marks.\label{fig:fig3}}
\end{figure*}

{\it Calibration Images}.  In order to improve observing efficiency,
it was decided not to attempt photometric calibration of the direct images
on a nightly basis.  The key factor driving this decision was the desire
to be able to utilize clear but not strictly photometric nights for
data acquisition.  Hence we developed a scheme of post-calibration, where on
just a handful of photometric nights we observed a portion of
each field using only the central 512 $\times$ 512 pixels of the CCD.
Exposure times were short, the small format images read out in seconds,
and accurate positioning was not required.  Hence, we could observe a
large number of fields in a very short amount of time, interspersing
standard star observations throughout.  All photometric calibrations
were done using standards from Landolt (1992).  During the 
processing of the survey data, the fluxes in the brightest stars in these 
calibration images were then compared with the fluxes for the same stars 
in the much deeper survey images, yielding an accurate post-calibration 
of the main survey images.  These calibration images were often obtained 
during bright time, so that all scheduled dark time could be used for 
survey work.

\subsection{Observing Procedures}

We utilized a standardized observing procedure for all data taking.
Initial field acquisition was done using setting circles only, since
the telescope does not have position encoders.  A short test exposure
was taken, and the positioning improved by comparing the test image
to finder charts (photographic enlargements of the POSS).  This was
accomplished by moving the telescope by small increments using images 
displayed on a monitor that are taken with a video camera which is part 
of the offset-guider system.  By careful adjustment, a pointing accuracy 
of 0.5 to 1 arcmin was achieved.  This precision was adequate for our
purposes.

Standardized exposure times were used throughout.  These times were
established early in the project, and reflect a trade-off between our
combined goal of good depth and large areal coverage.  For observations 
made in direct mode that were taken early in the project, each field was 
imaged twice through the V filter for 360 s, and once through the B 
filter for 720 s.  After it was realized that the direct images were
substantially deeper than the spectral images (by more than 2 magnitudes),
the exposure times for the direct images were reduced to 300 s in V and
600 s in B.  The spectral observations consisted of four exposures of 720 s 
each.  Under non-optimal observing conditions (e.g., a slight haze present, 
the moon up), these exposure times were increased slightly to compensate,  
although in general very few images that have been used to construct the 
survey were taken under these lower-quality conditions.  
For example, we took almost no survey data while the moon was up, and then 
only when it was during crescent phase and/or very close to the horizon.

The multiple exposures were acquired in order to allow for the easy 
removal/rejection of particle events (cosmic rays) in the images.  In 
addition, the telescope was dithered between exposures by $\sim$10-15
arcsec in order to reduce the effect of lower quality regions on the
detector (e.g., bad columns, lower sensitivity pixels). 

Special care was taken to align the dispersion direction of the prism
with the CCD columns when observing in spectroscopic mode.  This was
accomplished by rotating the prism in its cell until good alignment
was achieved.  Although not strictly necessary, this step greatly
simplifies the procedure of extracting the spectra (see section 3.2).

Calibration data acquired on each night of observing included bias
images (zeros) and twilight flats.  The twilight flats proved adequate 
for our purposes, since our program requires flat backgrounds over modest 
areas of the CCD, but low-level gradients over larger scales could be 
tolerated.  The twilight flats typically produced reduced images that had 
edge-to-edge gradients of no more than $\sim$1\%, even for data taken 
through the prism.  We took five twilight flats through each filter, 
dithering the telescope between exposures.  Dark images were taken 
typically once per run, but as there was no appreciable dark level 
associated with this CCD, no correction for dark counts was applied.

\section{Image Processing}

\subsection{Preliminary Reductions}

As with any major observational survey, image processing and analysis are
major components of KISS.  One of the primary goals of the KISS project,
established during our preliminary discussions about the survey, is to 
carry out the bulk of the searching and parameter determination for the 
emission-line galaxy candidates in an automated fashion.  To that end, a 
substantial amount of new software was developed by members of the KISS 
collaboration.  The image processing packages adopted for our work were 
IRAF\footnote{IRAF is distributed by the National Optical Astronomy 
Observatories, which are operated by AURA, Inc.\ under cooperative agreement
with the National Science Foundation.} (for the US team) and MIDAS\footnote{MIDAS 
is an acronym for the European Southern Observatory software package -- Munich 
Image Data Analysis System.} (for the Russian team).  Most of the survey 
processing and analysis reported here has been done using the IRAF-based 
software, a complete description of which is given in Herrero \etal (2000).  
The MIDAS software package is described in Kniazev (1997) and Kniazev \etal 
(1998).  In this section we give an overview of the processing steps carried 
out on the data.

The preliminary processing of the images follow fairly standard procedures.
Our first step is to flag the saturated pixels in each image, in order to
be able to track them through subsequent analysis steps.  This is accomplished
by inverting the counts in the saturated pixels (setting pixels with fluxes at
32767 to -32767).  Next the bias is removed from each image using a combination
of the overscan region (for the mean bias level) and an average of ten bias
images (for the 2D bias structure).  As mentioned above, no significant dark 
level is present in the images produced by the CCD, so no dark correction is
carried out.  The individual twilight flats are combined (using a median
combine algorithm) into a single flat-field image, which is then applied to
the survey images.  Finally, the several bad columns in the S2KA chip are
interpolated over to remove the worst of the cosmetic defects of the CCD.

The next stage in the processing involves aligning and combining the individual
images of each field into a single, deep combined image.  The same procedures
are followed for the spectral and direct images.  An IRAF script was written 
as a front-end to the IRAF task {\tt imalign} to 
expedite the process.  This script displays each image in the group in turn,
and allows the user to mark stars used to create the rough offsets between the
images that are used by {\tt imalign}.  Once the images are shifted to a
common center, they are combined to create a single, deep image of each field
which is fairly free of particle events (although a few invariably survive 
the process).  After the end of the alignment and combining stage, the KISS
data consist of a single combined spectral image, a deep combined direct
image (created from the two V and one B direct images), a combined V direct
image, the single B direct image, plus the small-format calibration images.  
All three direct images have precisely the same alignment.

Prior to processing the data through the KISS software package, a number of
steps are carried out on the images in order to streamline the analysis.
These steps are referred to as the pre-KISS processing stage,
and amount to converting the data into a uniform format and ensuring that
all information necessary for the subsequent KISS processing is present
in the image headers.  The steps include (1) rotating and/or flipping
the images to a standard orientation (N up, E left); (2) adding certain
header keywords that are not placed in the image headers by the acquisition
software; (3) computing and inserting into the headers the sidereal time
and airmass for the time of the observation; (4) determining accurate central 
coordinates and field rotation values; (5) measuring the mean FWHM of the 
stellar point-spread function (PSF) and writing it into the header.  For 
many of the steps, special IRAF scripts have been written to streamline the 
process.  For example, step 4 is carried out using a script which displays an 
image, overlays the location of all stars in the Guide Star Catalog (GSC, 
Lasker \etal 1990), and allows the user to interactively match objects within 
the image to their corresponding GSC marker and hence determine the coordinate 
offset between the field center and the less accurate telescope coordinates.
A second stage of the program allows the user to measure and correct for any 
field rotation present (usually this was no more than 1--2$\arcdeg$).

\subsection{KISS Package Processing}

All subsequent processing and analysis of the KISS data occurs within
the KISS software package.  This suite of scripts and programs is
described in detail in Herrero \etal (2000).  Here we give a brief
summary of the tasks carried out on the data.

The goal of the KISS processing is to extract all useful information
from both the direct and spectral images and store it in a conveniently
accessible format.  For the latter, we adopted the STSDAS Tables
format.  The first task of the KISS package is therefore the creation
of an empty KISS table for each survey field.  All relevant 
information from the image headers is transferred into the table
header at this step.  Subsequently, all work by the KISS package
programs modifies the data contained in the KISS tables, not the images.
The processing steps can be conveniently described in terms of seven
discrete functions, or modules: (1) object detection and inventory;
(2) photometry and object classification; (3) astrometry; (4) spectral
image coordinate mapping and background subtraction; (5) spectral
extraction and overlap correction; (6) emission line detection;
(7) spectral parameter measurement. 

\noindent(1) {\it Object Detection and Inventory}.  The goal of this
module is to locate all objects in the combined direct image with peak
fluxes greater than 5 times the local RMS noise (background plus 
instrumental noise).  A series of routines provide for the automatic
detection and cataloging of all such objects in each field, plus the
facility to interactively reject spurious sources (caused, for example,
by diffraction spikes and rings around bright stars) and add ones missed
by the automated software.  Great care is taken during this step to
ensure a high degree of reliability in the final list of objects in the
KISS database for each field.  For a typical field, all objects brighter
than V $\approx$ 20.5 are included.  At the end of this stage of processing,
the KISS table includes an entry for each detected object.  The key
information recorded for each object includes an estimate of its brightness 
and an accurate pixel position (x,y) in the image.  An average of 5200 
objects are found per field.  However, there is a large variation in the 
number of detected objects within each field, ranging from 3000 to 12000.
Most of this variation is due to changing Galactic latitude and longitude.

\noindent(2) {\it Photometry and Classification}.  In this second
module, the V and B filter images are used along with the small-format
calibration images to derive accurate magnitudes and B$-$V colors for
all objects in each field.  The multi-aperture photometry is also used
to classify the objects brighter than V = 19 as either stars or galaxies 
(see \cite{kearns00}).  Objects 
classified as galaxies are then remeasured using apertures tailored for 
extended objects.  As part of the photometry process, an inventory of 
nearby sources is created for each object, and those whose photometry 
is potentially contaminated by near neighbors are flagged in the table.  
At the end of this stage, each object in the KISS table possesses 
calibrated B and V magnitudes, a B$-$V color, and, for the brighter 
objects, a classification type.  Formal errors for all photometric 
values are also tabulated.

\noindent(3) {\it Astrometry}.  Accurate positions for all objects in the 
KISS table are determined.  Stars contained in the GSC are automatically 
located in each survey field, and a full astrometric solution is obtained 
and applied to all of the objects in that field.  An average of roughly 150 
GSC stars are used per field in the solution, and the formal RMS errors in 
the final positions are typically 0.20--0.25 arcsec in both right ascension 
and declination.

\noindent(4) {\it Spectral Coordinate Mapping \& Background Subtraction}.  
Once all of the relevant information is gleaned from the direct images, the
processing steps begin to focus on the spectral images.  The first step 
involves establishing the coordinate transformation between pixel positions
in the direct images and the corresponding positions in objective-prism
images.  Since the direct images are used exclusively for object 
identification, we must be able to locate accurately the
position of each object in the spectral images based on its direct image
coordinates.  This is accomplished by an interactive routine that requires
the user to provide an initial identification of several objects in both the
direct and spectral image.  This initial input provides the software with a 
coarse offset between the two fields, after which many additional stars are 
selected automatically, based on brightness, to be used to derive a full
two-dimensional coordinate transformation between the two images.  This
includes terms for the x-y offsets, rotation, and magnification (stretching).
Once the transformation is derived, the x-y positions in the spectral
image of each object in the KISS table are computed and recorded.  Objects
which are not completely contained within both the direct and spectral
images (i.e., lie entirely or partially off of the edge of one of the two
images) are then flagged and removed from the sample.  In a subsequent
step, the sky background in the spectral image is determined and removed,
using a routine that masks each spectrum, then uses the remaining pixels
to generate a smoothed, median-filtered background image.

\noindent(5) {\it Spectral Extraction and Overlap Correction}.  This module
is at the core of the KISS processing.  All objects present in the KISS table
have their objective-prism spectra extracted into a one-dimensional (1D) format
by summing the flux perpendicular to the dispersion direction.  Experiments
suggested that optimal signal-to-noise is achieved by summing over 4 pixels.
The resulting 1D spectra are written to the KISS database table for each
object.  As part of the extraction process, any spectrum that is overlapped
by another is flagged as such by the software, and an algorithm is then applied
which attempts to correct for this overlap.  The correction uses template
2D spectra generated from isolated bright stars in each field, and makes
use of the relative separations and brightnesses of the objects {\it as
measured from the direct images} to determine the proper correction on
a pixel-by-pixel basis.  This technique successfully corrects for the effects
of overlapping spectra in all but the most extreme cases (i.e., those with
large brightness differences between the two overlapping objects).

All processing up to this stage is carried out on all objects in each field 
in a uniform manner.  Nothing that occurs prior to this point is specific to 
the task of detecting emission-line objects.  Only at this point, when the 
KISS table contains complete information about the photometric and astrometric 
characteristics of each object, as well as the extracted 1D spectra,
does the KISS software become specialized to the task of finding ELGs.

\noindent(6) {\it Emission Line Detection}. This module centers upon a
non-interactive routine which examines each spectrum for the presence of a
strong emission feature.  The program fits a low-order
polynomial to a median-smoothed version of each spectrum, subtracts the
resulting fit from the original spectrum, and then searches for pixels with 
a flux level above a user-specified threshold.  Experimentation with the 
actual data led us to adopt as our threshold a level of 5 times the RMS 
noise in each spectrum.  Thus, one should be able to place a high degree 
of reliability on objects cataloged as ELG candidates.  Once the program 
completes the automated search, all
selected candidates are examined visually, both in the the objective-prism
images and in the extracted spectra.  This process usually leads to the
rejection of many false detections.  The average number of objects found
in each field for the red spectral data was 80, of which only 25--30\%\ 
turned out to be valid ELG candidates.  Many of the remaining objects
were found to be objects for which the overlap correction was imperfect
(e.g., when a very bright object overlaps a faint one), or residual cosmic rays
not rejected during the combining step.  Each object that passes the
inspection stage is flagged in the KISS table as being an ELG candidate.

\noindent(7) {\it Measuring the Spectra}.  The final step in the processing
involves measuring the spectrum of each of the ELG candidates.  A rough wavelength
and flux scale is assigned to the objects in each field, and an interactive
program displays the spectrum of each ELG candidate and allows the user
to measure the putative emission line for both position and total strength.  
The former is used to estimate the redshift of the galaxy, the latter to estimate
the line flux and equivalent width.  Although only relatively crude measurements,
these quantities provide extremely useful information regarding the survey 
objects.  The accuracy and utility of these measurements is discussed 
in the following section.

\section{Results}

\par The first two lists of KISS ELGs are presented in Salzer et al (2000a,
hereafter KB1) for the blue ([\ion{O}{3}]-selected) survey and in Salzer et al 
(2000b, hereafter KR1) for the red (H$\alpha$-selected) survey.  Here, we 
summarize the main results obtained from these first two survey strips,
and give general properties of the survey data.  The reader is referred to 
the two survey papers for complete details regarding the make-up of the
ELG candidate lists, the properties of the ELGs, and the estimated completeness
of the individual ELG catalogs.

\par KB1 covers 102 Schmidt fields, aligned in a strip of constant declination
($\delta$(1950) = 29$\arcdeg$ 30$\arcmin$).  The R.A. range covered is 8$^h$ 
30$^m$ to 17$^h$ 0$^m$.  This area was chosen to overlap completely the Century 
Redshift Survey (Geller \etal 1997), which provides photometric and redshift 
information for a sample of 1762 field galaxies complete to m$_R$ = 16.13.  
After accounting for the overlap between the survey fields, the total area 
covered by KB1 is 116.6 square degrees.   A total of 223 ELG candidates were 
selected for inclusion in the first blue list, which results in a surface 
density of 1.91 ELGs per square degree.

\par The areal coverage of KR1 is substantially less than that of KB1,
only 54 Schmidt fields.  The red spectral data were taken to overlap with
the existing KB1 direct images, so the two surveys cover essentially the
same area where they overlap.  The R.A. range of KR1 is from 12$^h$ 15$^m$
to 17$^h$ 0$^m$, excepting the three survey fields between 14$^h$ 30$^m$ and
14$^h$ 45$^m$.  The total area covered by KR1 is 62.2 sqaure degrees,
and a total of 1128 ELG candidates are included in this list.  The surface
density of H$\alpha$-selected KISS objects is 18.14 ELGs per square degree!
For the area in common between KB1 and KR1, there are 125 blue-selected ELG 
candidates.  Of these, 109 (87\%) are also cataloged in KR1, and several of
the remaining 16 KB1 candidates not included in KR1 are detected in the red
but at levels slightly below our selection threshold.  For this reason, 
coupled with the substantially higher detection rate of the red-selected 
spectra, all subsequent survey work will be undertaken using red spectra only.

Examples of several newly discovered ELGs are shown in Figures 4 and 5.  Figure 4
displays three small sections of a single survey field, showing four KISS ELGs.
The objects chosen for this illustration display a range of observational
characteristics which reflects the variety seen in the overall survey.  These
galaxies show a substantial range in brightness, emission-line strength and
contrast (equivalent width), and redshift (as indicated by line position).
Figure 5 shows plots of the extracted spectra for all four ELGs.

\begin{figure*}
\plotone{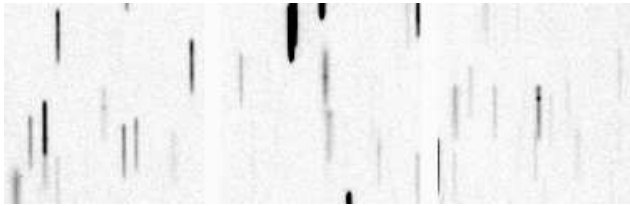}
\figcaption[fig4.eps]{Examples of 4 newly discovered ELGs detected in the KISS survey.  
{\it Left:} An example of a fairly faint (B=18.6) ELG. {\it Center:} Two
ELG candidates: a lower-redshift object just above center, and a high-redshift 
galaxy just below it. {\it Right:} An intermediate-redshift object with a 
fairly strong emission line.\label{fig:fig4}}
\end{figure*}
\begin{figure*}
\plotone{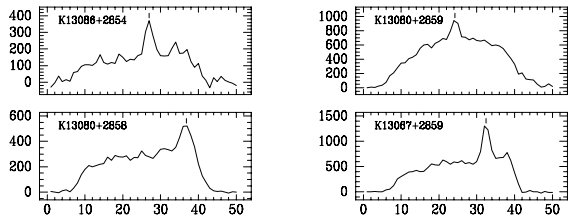}
\figcaption[fig5.eps]{Plots of the extracted spectra for the four ELGs displayed in Figure 
\ref{fig:fig4}.  K13086+2854 corresponds to the left image, K13080+2859 (upper)
and K13080+2858 (lower) to the center image, and K13067+2859 to the right.
Each spectral plot covers $\sim$800 \AA, in the range 6400 -- 7200 \AA.
Lines mark the location of the H$\alpha$ emission line.\label{fig:fig5}}
\end{figure*}

A summary of the previous Schmidt surveys for active galaxies is given
in Table 1.  This table is not meant to be complete, but rather
to give examples of existing surveys for comparison purposes.  The surveys
listed are grouped by the type of object detection method employed 
(column 2): color selection, UV-excess selection, emission-line selection, 
or a combination thereof.  Schmidt/objective-prism surveys for QSOs (e.g.,
PG Survey) are {\bf not} included or discussed here; numbers in the tables 
for surveys that cataloged both ELGs and QSOs (e.g., UM, Case, Kiso) represent 
only the galaxies.

All of these surveys, with the exception of KISS, share an important common 
feature: they were carried out using photographic plates.  Even in the era of
high quantum efficiency CCD detectors, the large areal coverage of
photographic plates has made them the detector of choice for most Schmidt
surveys.  However, the relatively low sensitivity of the plates has limited
the depths of the blue/emission-line galaxy surveys, particularly those
which employ an objective prism, since dispersing the light makes the
threshold effect of photographic emulsions even more severe for faint 
objects. 

A few of the surveys listed in Table 1 have utilized scanners to
digitize the survey plates, and then used automated detection software to
search for emission-line galaxies.  These include the Marseille (Surace
\& Comte 1998) and Hamburg (Popescu \etal 1996) surveys, as well as the 
most recent installment of the UCM survey (Alonso \etal 1999).  An 
additional survey of this type that is currently in progress is the 
Hamburg/SAO survey (Lipovetsky \etal 1998, Ugryumov \etal 1999), which is 
similar in most respects to the Hamburg survey.

Two key figures of merit for these surveys are the surface density of objects 
detected and the completeness limit (or depth).  Since objects in a line-selected
sample are chosen from the survey images via a complicated combination of
emission-line flux and equivalent width (\cite{SSG86}, \cite{GO87}, \cite{S89}), 
it is not possible to
compare the depths of these surveys directly with those of the color and 
UV-excess selected samples, which are effectively magnitude-limited samples 
(albeit with complex selection functions).  {\it Line-selected surveys can not
be treated as magnitude-limited samples}.  For comparison purposes only, we 
include in column (6) of the table the {\it median blue apparent magnitude} (when 
available) for some of the line-selected surveys.  This should not be taken as 
anything more than an illustrative number, but it at least allows the various
types of surveys to be compared in a rough sense.  We stress that one cannot 
infer accurate relative depths of the various surveys using the numbers in column 
(6).  Still, one does get a feel for the survey depths from these numbers.

The color and UV-excess surveys are all relatively shallow, being complete to 
only m$_B$ of between 14.8 and 16.0.  Some of the line-selected samples appear 
to be sensitive to much fainter galaxies.  The UM survey has a median m$_B$ = 16.9,
SBS has a median m$_B$ of roughly 17.0, while for the H$\alpha$-selected UCM survey,
a median blue magnitude of $\sim$16.5 is estimated.  By comparison, the {\it median 
m$_B$ for the H$\alpha$-selected KISS galaxies is 18.1}, substantially fainter 
than the other existing surveys.

Since it is not possible to make a fair comparison between surveys based on
their apparent magnitudes, a more relevant value is the surface density of
objects.  Among the older surveys, those with the highest numbers of
objects include the Kiso, SBS, and Case surveys.   Kiso is purely color-selected,
while the other two use a hybrid selection method, considering both UV-excess and
emission lines.  Follow-up work on the Case survey (\cite{S95}) indicates that, 
to first order, it can be treated like a color-selected survey, which is why its
completeness limit is listed directly.  The previous line-selected surveys with the
highest surface densities are the UM and UCM surveys, both with roughly one ELG
for every two square degrees surveyed.  Both of the KISS samples detect objects
at a substantially higher rate.  The blue, [\ion{O}{3}]-selected survey has a 
density four times higher than the UM survey, with which it compares most directly, 
and 19 times higher than the famous Markarian survey.  The numbers for the red 
(H$\alpha$-selected) KISS sample are even more impressive: a factor of 32 times
higher surface density than the UCM survey, and 181 times higher than Markarian!

An additional survey that has a number of important similarities to KISS, but that
is missing from Table 1, is the Palomar Transit Grism Survey (PTGS, \cite{SSG94}).
The reason PTGS is not included in the table is the fact that it is not
an objective-prism survey.  With that distinction aside, however, PTGS and KISS
share a number of attributes.  PTGS employed CCDs as their detectors, and
hence reached to quite faint magnitudes.  Although PTGS was designed primarily
to detect high-redshift QSOs, a large fraction of the objects discovered were
in fact low-to-moderate redshift galaxies with strong emission lines.  The
surface density of objects discovered in the two surveys is comparable.  Some 
of the software algorithms developed for KISS were patterned after those first 
used by PTGS.  In particular, we adopted a line-detection algorithm very similar 
to that used by PTGS (see \cite{herrero00}).

As already mentioned, KISS is the first large-scale survey to utilize CCDs for 
objective-prism spectroscopy.  Other groups have also begun pursuing digital
objective-prism survey observations using techniques similar to KISS.  These
include the QUEST (QUasar Equatorial Survey Team; Sabbey 1999) survey and the 
UCM-CIDA survey (Gallego \etal 1999).  

In the following subsections, we illustrate various properties of the 
KISS data, including the depth and photometric quality of the direct images, 
the quality of our positions, and the reliability of the redshift and 
line-strength information recorded in the survey tables.

\subsection{Photometric Quality}

The photometric properties of the survey data are conveniently
summarized in Figure 6, which shows the data for all objects contained
within a single KISS field (F1455).  A total of 10,078 objects were
detected in this field.  The upper panel of the figure shows the
histogram of the measured V magnitudes, which peaks at V $\approx$ 21.0 
(B $\approx$ 22.0) and then drops off rapidly at fainter magnitudes.  
It is likely that many of the very faintest objects found by the software 
(V $>$ 22.5) are spurious sources.  These faint objects play no role in 
the ELG survey, since we do not detect sources this faint in the 
spectral images.  The lower panel in Figure 6 shows the histogram of
B$-$V color for those sources with reasonably reliable photometry 
(those with V $<$ 21.5).  The median color is 0.98; clearly 
this sample of objects is dominated by late type stars.  
The need for automated software to carry out the survey is clear from
these plots: it would be an impossible chore to extract the 15 -- 30
ELG candidates from among the 3,000 to 12,000 objects in each field
using eye searches alone.

\begin{figure*}[ht]
\plotone{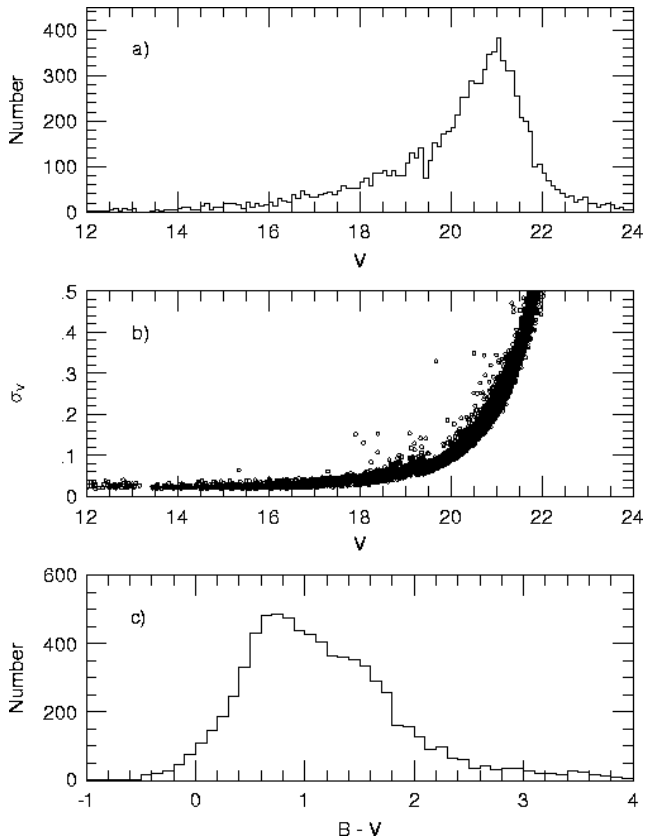}
\figcaption[fig6.eps]{(a) Histogram of the V magnitudes for all 10078 objects
detected in survey field F1455.  The number of objects detected peaks at
V = 21.1.  (b) The formal photometric error (in magnitudes) for all
objects detected in field F1455, plotted as a function of V magnitude.
(c) Histogram of B$-$V color for all objects in field F1455 with V $<$ 21.5.
\label{fig:fig6}}
\end{figure*}

Figure 6b shows the photometric errors associated with the
V magnitudes in field F1455.  At bright magnitudes, the errors are
dominated by uncertainties in the photometric calibration, which are
somewhat higher than one might expect due to the
``bootstrap" nature of our calibration scheme.  The formal errors in
our nightly zero-point constants are typically better than 0.02
magnitude.  However, the uncertainty in the photometric offset 
between the deep survey images and the short-exposure calibration 
images is usually at about the same level.  Hence, for bright objects
our photometric precision is typically $\pm$0.03-0.04 magnitude.  As
fainter magnitudes are approached, the sky noise becomes an increasingly
important source of uncertainty.  To compensate partially for this,
fainter objects are measured through progressively smaller apertures,
and an aperture correction is applied to the magnitudes.  The smallest
aperture used is 8 arcsec in diameter, which is used for objects with
V $>$ 19.5.  Beyond this magnitude, the photometric errors increase
sharply.  For field F1455, a characteristic value of $\sigma_V$ is 0.10 magnitude
at V = 20.0, 0.23 magnitude at V = 21.0, and 0.36 magnitude at V = 21.5.

	The depth of the direct images are fairly uniform for the full
survey.  One way to assess the depth of the imaging data for each field
is to measure the peak in the apparent magnitude histogram.  When this
is done for all 102 fields of KB1, the mean value of V$_{peak}$ = 20.5 is
determined.  This is between one and two magnitudes deeper than the 
spectral images.  The scatter about the mean is 0.25 magnitude, with 
only four fields having V$_{peak}$ slightly brighter than 20.0.  There is
no trend in the variation of V$_{peak}$ with position on the sky.

	It was difficult to find adequate comparison sources for use in
evaluating the quality of the KISS photometry.  The survey data yield
high quality photometry in the magnitude range V = 13 to 20.  Brighter
than V = 12 to 13, stars are saturated in our images.  At magnitudes fainter
than our saturation limit, most available photometric data are photographic, 
and hence not terribly useful in terms of providing a critical test of our 
precision.  One list of fainter stars with published photoelectric photometry 
which overlaps our survey area is given in Purgathofer (1969).  This list of 
33 stars is contained within our F1305 survey field, and includes stars with 
a brightness range of V = 12.8 to 18.5.  Comparison of the KISS photometry 
with that published by Purgathofer shows good agreement.  The mean offset 
in V between the two datasets is -0.001, with a scatter of 0.039 magnitude.
This level of scatter is fully consistent with the formal errors in the
KISS photometry.  Similar agreement is found for the B photometry,
although the mean offset between the two data sets of $\Delta$B = 0.043
is marginally significant.  Further work will be required to better
assess the reality of this offset.  It would appear, however, that there 
are no large ($>$ 0.05 magnitude) photometric offsets in our data, at 
least for magnitudes brighter than 18.5.

\subsection{Astrometric Quality}

	A goal of the project is to provide sub-arcsecond precision
in our astrometry.  This motivation is driven in a large part by our
expectation that follow-up spectra would be most efficiently obtained
using multi-fiber spectrographs (e.g., Hydra at Kitt Peak), and accurate
positions are a prerequisite for such instruments.  For this reason, a 
full astrometric solution is obtained for each survey field, using the
Guide Star Catalog (Lasker \etal 1990) as our source for positions. 

	An internal estimate of our positional accuracy is provided
by our own astrometry routine, which after computing the astrometric
solution for a given field compares the final calculated positions of 
the GSC stars with their cataloged values.  The average value of the RMS 
scatter between the calculated and catalog positions for the 102 KISS 
fields is 0.25 arcsec in R.A. and 0.20 arcsec in declination.  Given our
image scale of 2.03 arcsec/pixel, we are satisfied with this level
of precision.  These values represent the typical {\it relative} positional
accuracy for objects within a given field, a number which is relevant
when considering the positional precision that is necessary when using 
a multi-fiber spectrograph. 

	The estimate of the absolute precision of our positions was
obtained by comparing our astrometry with that of Willmer \etal (1996), 
who performed detailed astrometric solutions for photographic plates as part
of a deep survey for galaxies in the North Galactic Pole.  Four of the
KISS fields overlap parts of the Willmer \etal survey.  We matched the 
stellar sources in Willmer \etal with their corresponding objects in
the KISS database, and computed the mean positional offsets between the
two catalogs for the 695 stars in common.  The range of magnitudes 
covered by the stars in this overlap sample was 11.0 -- 15.5.  The average 
difference in R.A. (in the sense KISS $-$ Willmer) was 0.71 arcsec, with
individual fields having offsets as large as 1.06 arcsec and as small
as 0.35 arcsec.  Positional agreement in declination was somewhat
better, with an average difference (again, KISS $-$ Willmer) of 0.29 arcsec
and a range of 0.07 to 0.46 for individual fields.  The differences
between the two samples are most likely due to the different sources
of astrometric materials used by the two surveys.  Willmer \etal used
an astrometric catalog from the Lick Astrograph, and showed that there
are irregularities with the GSC at the level found here.  We conclude 
that the {\it absolute} astrometric precision of
KISS is limited by the precision of the GSC, but is probably better than 
1 arcsec for most fields.  There is no indication of errors as large as 
2 arcsec.  However, we point out that this detailed comparison was 
carried out over a limited area of the overall survey.

\subsection{Morphology}

	The KISS direct images have a rather coarse scale of 2.03 
arcsec/pixel.  This makes detailed morphological analysis difficult
for objects fainter than V $\approx$ 19.  For brighter objects, the
KISS software package provides automated object type classification.  
At this time, the classification is limited to a simple star vs. galaxy 
designation.  The distinction between the
two classes of objects is important, however, since various KISS
routines make use of the object type classification in order to provide
better results.  For example, those objects flagged as galaxies are
rephotometered with appropriately sized apertures, since the main 
photometry routine uses apertures tuned to stellar sources.  The 
emission-line detection routine 
also makes use of the object classification to help eliminate the
many false detections that occur for late type stars.  Stars of spectral
type M have objective-prism spectra which mimic those of ELGs in the
red spectral region.  Our software automatically rejects objects
classified as stars that have colors redder than B$-$V = 1.30.

	The automated classification has the added advantage that we
can use the KISS imaging data to produce a digital galaxy catalog for
the area covered by the survey.  This catalog includes accurate B and V
photometry to V = 19.0.  Kearns \etal (2000) details the use of the
object classification software to produce such a galaxy catalog, and
then uses the catalog to do a number count analysis (see Section 5).

\subsection{Confirmation of ELG Status}

	Beginning in 1998, follow-up spectra of candidate KISS ELGs were
obtained using both multi-fiber and single-slit spectrographs.  During our
first two seasons of spectroscopic follow-up, 428 candidates from KR1 were
observed.  Of these, 402 (94\%) were confirmed to be {\it bona fide} 
ELGs.  Of the 26 unconfirmed KISS objects,  16 were found to be Galactic
stars, while 10 were galaxies without significant emission.  Inspection 
of the original objective-prism data for these unconfirmed objects
reveals that they were in all cases ``dubious" candidates, in the sense
that the apparent emission line was questionable.  In all cases, the
putative line turned out to be either noise (e.g., a remnant cosmic ray)
or an artifact of two overlapping spectra.  The results of our
follow-up spectra will be valuable in helping to fine tune the selection
process for subsequent survey lists.

	A complete description of the follow-up spectroscopy obtained
to date for the KISS ELG sample is given in Gronwall \etal (2000b).

\subsection{Redshift Comparison}

	One of the parameters we measure directly from the survey data
is the redshift of each ELG.  The assumption made is that the line observed
in the objective-prism spectrum is either H$\alpha$ (for red spectra) or
[\ion{O}{3}] (for blue spectra).  To date, only ELG candidates from KR1 have 
been observed spectroscopically, so the results discussed here and in the 
following subsection refer only to the red survey, and the detected line 
is assumed to be H$\alpha$.  This is usually a good assumption: of the 402
confirmed ELGs from the follow-up spectra mentioned above, only eight (2\%) are
high redshift objects where [\ion{O}{3}], H$\beta$, or, in one case, H$\gamma$,
is shifted into the spectral region covered by our red spectra.   Seven of the
eight are AGNs, with redshifts between 0.328 and 0.550.  In all other cases, 
the observed emission line was in fact H$\alpha$.  Given this result, we were 
interested in determining the level of precision in our redshifts estimated from 
the survey data.

	Figure 7 shows a direct comparison of survey and slit-spectra
redshifts for all objects with follow-up spectra (excluding the
high-z objects).  A very tight, linear correlation is seen.  For redshifts
up to z $\approx$ 0.07, the KISS redshift is seen to be an excellent
predictor of the true redshift.  The RMS scatter about the line z$_{KISS}$
= z$_{slit}$ is only 0.0028 (830 \kms) for z$_{slit}$ $<$ 0.07.  This number
is consistent with expectations based on the value of the dispersion provided
by the prism.  We consider this level of redshift precision to be excellent, 
and allows us to use the survey data themselves for certain types of analyses 
(see Section 5).

\begin{figure*}[ht]
\plotone{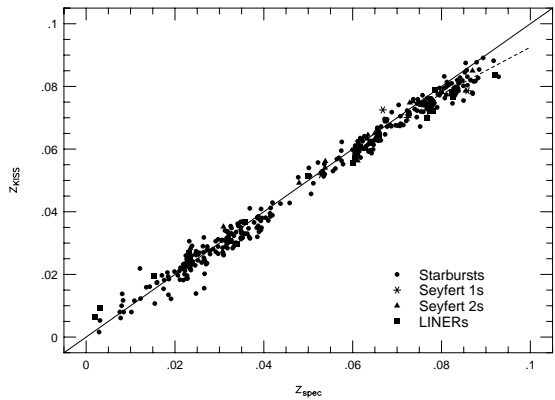}
\figcaption[fig7.eps]{Redshifts determined from the objective-prism spectra are 
compared to redshifts determined from follow-up spectra for 394 KISS ELGs.
Galaxies of different types are plotted with the symbols shown in the legend.
The solid line indicates z$_{KISS}$ = z$_{spec}$; it is not a fit.  The dashed
line is a fit to the galaxies with z$_{spec}$ $>$ 0.07, where the onset of the
survey filter cutoff results in a systematic underestimation of the redshift
by the KISS spectra.
\label{fig:fig7}}
\end{figure*}

	At higher redshifts, z$_{KISS}$ is seen to underestimate 
the true redshift by a modest amount.  This is due to the 
steep cut-off of the survey filter.  As the H$\alpha$ line redshifts into 
the wavelength region were the filter transmission drops off sharply, the 
red side of the emission line is reduced in strength, which pushes the 
flux-weighted line centroid to lower wavelengths (and hence lower redshifts).
A correction for this effect can be obtained by fitting the residuals in
Figure 7 at z$_{slit}$ $>$ 0.07 and applying the resulting correction to
the high-z portion of the KISS sample.  An example of a linear correction 
fit is shown as a dashed line in the figure.

	One additional comparison that we can make in terms of the redshift
precision is with KISS galaxies which are also members of the UCM survey.
Comparison of the KR1 list with the UCM objects observed by Gallego \etal
(1996) shows that there are 17 objects in common.  The mean redshift 
difference for these galaxies (in the sense z$_{UCM}$ $-$ z$_{KISS}$) is
$\Delta$z = -0.00012 (-35 \kms), while the RMS scatter is 0.0030 (900 \kms).  
Once again, we see that the objective-prism redshifts provide reasonable 
estimates of the true redshifts, although of limited precision.

\subsection{Line Flux and EW Comparison}

	The line fluxes and equivalent widths (EWs) measured directly from the 
objective-prism spectra are extremely valuable for establishing the completeness
limits of the survey.    
In addition, the H$\alpha$ fluxes from the red survey, if reliably calibrated, can 
be used to estimate the total star-formation rate of each KISS galaxy without the 
need for follow-up spectroscopy (e.g., Gronwall 1999). 

	The line fluxes are calibrated in a two-step process.  First, the spectra
from all of the survey fields are placed on the same flux scale by comparing the
observed fluxes from a sample of stars selected to fall within a narrow range of
color and apparent magnitude.  This procedure corrects the data from each field for
overall throughput variations, and ensures that the fluxes from the objective-prism 
data can be compared directly.  The absolute flux calibration of the H$\alpha$ line
is achieved using follow-up spectra obtained with slit spectrographs.  We found that 
spectra obtained through fiber-fed spectrographs systematically underestimate the true 
H$\alpha$ flux, hence we only used spectra obtained on photometric nights through modest
slit widths (typically 2 arcsec) to calibrate the objective-prism fluxes.  Details
of the flux calibration are given in KR1. 

	To assess the accuracy of our flux calibration using an independent dataset, 
we compare our objective-prism
measured line strengths against the slit spectra of Gallego \etal (1996) for the 
objects in common.  Since our objective-prism spectra are of such low dispersion, the 
[\ion{N}{2}]$\lambda\lambda$6548,6583 doublet is always blended with H$\alpha$.  Thus
any measured line flux and equivalent width is really H$\alpha$ plus [\ion{N}{2}].
For this reason, we plot in Figure 8 the Gallego \etal H$\alpha$ + [\ion{N}{2}] fluxes 
vs. the KISS objective-prism line fluxes.  There is extremely good agreement between 
the two datasets, although a few discrepant sources are evident.  Two of the deviant 
objects lie below the F$_{UCM}$ = F$_{KISS}$ line, suggesting that the flux measured by 
Gallego \etal may be underestimating the total H$\alpha$ emission for these objects.  
This would be the case if the emission were more extended than the width of the slit 
used by Gallego \etal, which appears to be true for both galaxies.  The deviant
point at the top of the plot most likely represents a case of the KISS flux being
underestimated.  This object (designated UCM 1259+2934) has a bright neighboring
galaxy, for which an overlap correction was applied (see Section 3.2 and 
\cite{herrero00}).  Presumably the objective-prism flux was inappropriately reduced 
by this correction.  In general, however, 
the KISS fluxes agree well with fluxes derived from slit spectra when the
emission region is unresolved.  For extended emission, the KISS spectra recover more
of the total emission than do slit (or fiber-fed) spectra.  Hence, despite being rather
coarse in nature, the KISS spectra may be the dataset of choice for use in deriving
parameters such as the total star-formation rate for this sample of galaxies.

\begin{figure*}[ht]
\plotone{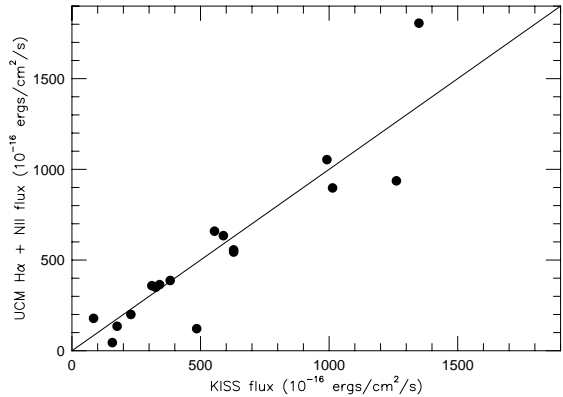}
\figcaption[fig8.eps]{Comparison of the H$\alpha$ + [\ion{N}{2}] line fluxes 
and KISS objective-prism fluxes for galaxies observed by Gallego \etal 1996
that are common to both the KISS and UCM surveys.  The solid line denotes
F$_{KISS}$ = F$_{UCM}$.
\label{fig:fig8}}
\end{figure*}

\section{Uses of the Survey}

KISS provides a very deep, statistically complete sample of active and star-forming
galaxies.  A huge advantage over most previous objective-prism surveys is that
the completeness limits and selection function of KISS can be quantified from the
survey data themselves.  Such a sample can be used to study a wide range of problems 
in modern astrophysics.  In this section we briefly discuss a number of possible
applications of the KISS sample.  The list can be broken up into four broad categories.

{\bf Galaxian Properties / Galaxy Evolution.}  The KISS galaxies should provide an
excellent catalog for addressing a number of issues regarding the properties of
star-forming and active galaxies, particularly the types of questions that require
statistically complete datasets.  For example, KISS should yield accurate numbers
for the frequency of AGN/starburst activity within the general galaxian population.
Once follow-up spectra have been obtained for large enough samples of the KISS ELGs,
studies of the characteristics of the large-scale star-formation processes occurring 
in galaxies, the chemical enrichment/evolution in galaxies (particularly dwarfs), 
and the verification/quantification of a metallicity--luminosity relationship for 
starbursting galaxies will all be possible.  

Another study that should yield fruitful results will be to correlate the KISS catalog 
of ELGs with surveys at other wavelengths, such as the radio, FIR, and X-ray.  
Since KISS is so deep, it will match up well with a number of recent surveys.  These
multi-wavelength studies of the KISS sample should lead to new insights into the nature 
of the activity present in these galaxies.  The first stages of a correlation between 
the FIRST radio survey (\cite{FRS}) and KISS (the FIRST--KISS project) is already 
underway (\cite{EB99}), and a complementary project involving the IRAS survey 
database is planned.

{\bf Activity in Galaxies.}   The KISS survey will provide a deep, complete sample
of AGN which will be a valuable complement to existing samples of radio and X-ray
selected Seyferts.  One clear advantage of KISS over previous objective-prism
surveys is that the H$\alpha$ selection method means that a more complete sample
of nearby AGN will be detected.  For example, the H$\alpha$-selected portion of 
KISS is sensitive to LINERs (\cite{Heck}), while the older [\ion{O}{3}]-selected 
samples such as UM, Tololo, and Case are not.  Also, the H$\alpha$-selected sample 
appears to detect a much larger fraction of Seyfert 2 galaxies than previous 
surveys which were carried out in the blue.  This is because Seyfert 2s are typically 
more heavily reddened than Seyfert 1 galaxies.
One of the interesting topics that could be investigated is the study of the active 
stages of galaxian evolution.  In particular, what fraction of the local galaxy 
population currently host an active nucleus?  Or, assuming that all large galaxies
contain a central black hole, what fraction of the time are the nuclei ``turned on"?
Other areas of interest, particularly once follow-up spectra are available, will
be the physics of AGN/QSO phenomena with an emphasis on testing the unified model, 
the environmental influences on activity, and similarities/differences between AGN 
selected at optical, radio, and X-ray wavelengths.

{\bf Observational Cosmology.}   The KISS survey has the potential to contribute 
a great deal in the area of observational cosmology.   For example, pre-selecting 
galaxies that exhibit strong optical emission lines is an efficient way to carry out 
redshift survey work to probe large-scale structure (e.g, \cite{S89}; \cite{R94}).  
The surface density of the survey is sufficiently high (e.g., comparable to the
Century Survey of Geller \etal (1997)) that KISS-selected galaxies could quickly
map out the spatial distribution of field galaxies to z $\sim$ 0.095 in the areas
surveyed.  Further, line-selected samples of galaxies, such as the KISS, Tololo, UM, and 
UCM surveys, tend to preferentially detect dwarf galaxies (\cite{S89}).  Studies of
the spatial distribution of dwarfs (e.g., \cite{S94}, \cite{P95}, \cite{Pop97},
\cite{TM99}, \cite{LSLR}), made 
possible with dwarf-rich samples such as these, will allow for the study of biasing 
in the spatial distribution of low-mass galaxies.  Our preliminary analysis indicates 
a significant population of dwarf galaxies within nearby voids.  A detailed discussion 
of both of these issues appears in Lee, Salzer \& Gronwall (2000).

An important set of observables from the very early universe are the abundances of the
primordial elements (e.g., He, D, Li).  Key among these is helium.  By studying low
metallicity galaxies (i.e., those with relatively little chemical processing since
their formation), astronomers have tried to infer the primordial abundance of helium,
Y$_P$ (e.g., \cite{IT98}).  This has been an active area of research for some time,
yet there continues to be substantial debate over the correct value of Y$_P$.  The
effort to determine Y$_P$ accurately would be aided by the existence of a larger
sample of very metal poor galaxies.  However, despite focused efforts to discover 
ultra-low metallicity objects (Z $<$ Z$_\odot$/25, see \cite{KS}), very few such 
objects are known.  Ideal candidates for this work would possess strong emission lines,
to facilitate the abundance determination.  Further, they would tend to have weak
oxygen lines, due to their low metallicity.  The focus of previous searches on the
contents of the [\ion{O}{3}]-selected photographic surveys may then have created a huge
bias against finding more low-Z galaxies.  Selecting by H$\alpha$, as KISS does, may
well lead to the discovery of many previously unidentified low metallicity galaxies
with which to investigate the primordial helium abundance.

Finally, the KISS sample will be useful as a low-redshift comparison group to the 
galaxian population at higher redshift.  Understanding adequately the activity levels
seen in high-z samples of galaxies that are currently being studied using 10-meter
class ground-based telescopes and HST will require having a much better knowledge of
the galaxian population locally.  Because we can quantify the completeness limits of 
KISS, it will be an excellent sample to use for comparisons to higher redshift surveys.
We will be able to measure more accurately than ever before the space densities of 
starbursting galaxies locally, which may well have an impact on resolving the faint 
blue galaxy problem.  In addition, KISS will also provide a more accurate estimate of 
the local star-formation rate (SFR) density (e.g., \cite{JG}; \cite{Madau}) which,
when combined with high-redshift studies, will lead to a more precise understanding
of the evolution of the SFR as a function of cosmic epoch.  A preliminary 
effort to determine the local SFR density using the KISS galaxies is given in Gronwall 
(1999).  A more complete analysis is in progress (\cite{CG2}).

{\bf Surveys for Other Types of Objects.}  As stressed in Section 3.2, the processing and
analysis of the KISS data is completely general, up to the stage of searching for emission
lines.  Therefore, the information tabulated in the survey database files can easily
be utilized for surveys of other types of objects.  The only restriction is that the
type of object being searched for would need to have an identifiable spectral feature in
the bandpass of the KISS objective-prism spectra.  Although the restricted 
wavelength coverage of the KISS spectra limits the type of objects one can search for, 
there is nonetheless potential for these data beyond their primary purpose for KISS.

One such use that was explored was to employ the blue spectral data to look for
carbon stars (C stars).  The deep absorption band at $\lambda$ $\approx$ 5150 \AA\ 
was targeted.  The KISS package routine for detecting the emission lines in the
extracted spectra was replaced by a similar routine that examined the spectra of
red objects (B $-$ V $>$ 1.2) for the presence of a strong absorption feature at the
correct location.  A list of $\sim$75 C star candidates was generated for the 117
square degrees of the blue survey.  Spectroscopic follow-up observations have been
obtained for roughly a third of these candidates.  All three previously known C
stars in this area of the sky were recovered.  However, none of the other objects
for which we obtained higher dispersion spectra have turned out to be C stars.  
Rather, they are all late-type stars with spectral types of K and M.  Apparently,
the KISS spectra are not sensitive enough to substantially improve upon the existing
surveys for C stars.  Despite this result, we will continue to look for additional
opportunities to utilize the KISS data for the detection and cataloging of other
types of objects.

Another project that is currently underway, for the purpose of doing galaxian 
number counts, is the development of a digital catalog of galaxies that is 
complete to B $\approx$ 20.  While ultra-deep number counts have been
carried out with CCDs for years in small areas of the sky, the number counts at
brighter levels (B = 14 - 19) have either made use of photographic plates (and
suffered from questionable photometry) or used CCDs but covered only relatively
small areas (\simlt 10 deg$^2$).  Currently, KISS direct images exist for
over 200 deg$^2$ and possess well-calibrated photometry.  Our galaxy catalog
should provide some of the best available number count data in the magnitude
range indicated above, since it covers a large enough area to average over 
localized variations in the large-scale structure.  Kearns \etal (2000) 
presents the initial results of this project.  In addition, Kearns is also 
currently investigating the use of the KISS direct images to generate catalogs 
of low-surface-brightness galaxies.

\section{Summary}

We have presented a description of the KPNO International Spectroscopic
Survey (KISS), a new objective-prism survey for extragalactic emission-line 
objects.  KISS is the first large survey to combine CCD detectors with the
traditional Schmidt telescope slitless-spectroscopic-survey method, which
has proven to be extremely fruitful over the past three decades.  The 
superior sensitivity of the CCD allows us to survey to fainter flux levels
than were previously possible using photographic plates.  In addition,
the digital nature of the data will allow us to assess the selection
function and completeness limits of the survey far more accurately than
with photographic surveys.

KISS is being carried out on the Burrell Schmidt telescope.  The first
survey region was observed in both the red and blue spectral regions.
Both sets of objective-prism spectra were obtained over a restricted wavelength
region in order to minimize the problems of overlapping spectra and to reduce
the level of the sky background.  The blue spectral region covers 4800 - 5500 \AA,
and the primary emission line we select by is [\ion{O}{3}]$\lambda$5007.
The red spectra cover the wavelength range from 6400 - 7200 \AA, and objects
are detected primarily via the H$\alpha$ line.  Both the red and the blue
spectra detect galaxies out to z $\approx$ 0.095 via the primary line, and
are sensitive to higher-z galaxies over restricted redshift intervals
as other lines move into the survey bandpasses.

Because of the digital nature of the survey data, all of the candidate 
selection is carried out using automated software, although the final 
candidate lists are also checked visually.  In addition to the
objective-prism spectra used to select the ELG candidates, we also have direct
images taken through B and V filters which provide accurate astrometry,
photometry, and morphological information for all objects in each field.
These data, plus estimates of the redshift and line flux for each ELG
candidate obtained from the objective-prism spectra, yield a fairly
complete picture of each candidate galaxy without the need for additional
follow-up observations.  This makes the KISS database particularly
valuable for statistical studies of galaxian activity.

The results obtained from our first survey lists are extremely encouraging.
The first blue survey strip covers 117 deg$^2$ and includes 223 cataloged
ELG candidates.  With a surface density of just under 2 ELGs per square degree,
the blue KISS sample is substantially deeper than previous surveys of this type. 
The numbers for the H$\alpha$ (red) survey are even more impressive.  A
total of 1128 ELGs have been found in an area of 62 deg$^2$, for a surface
density of 18.1 ELGs per square degree.  This is 32 times the surface
density of the H$\alpha$-selected UCM survey, and 181 times that of the
UV-excess Markarian survey.

The first two survey lists will be presented in Salzer \etal (2000a,b),
along with a complete discussion of the characteristics of the two samples.
Additional survey lists will be published as the data are acquired, processed,
and cataloged.  To date, data covering 200 deg$^2$ of sky have been obtained,
and observations are continuing.  Our overall project goal is to cover in excess
of 300 deg$^2$ in a series of survey strips in selected areas of both
the north and south Galactic caps.  This will yield several thousand ELG 
candidates, which will be suitable for addressing a wide range of scientific
questions that cover nearly the full scope of extragalactic astronomy.

\acknowledgments

We gratefully acknowledge financial support for the KISS project through
NSF Presidential Faculty Award to JJS (NSF-AST-9553020), which was instrumental
in allowing for the international collaboration.  Additional support for
this project came from NSF grant AST-9616863 to TXT, and from Kitt Peak National
Observatory, which purchased the special filters used by KISS.  The following 
summer research students working at Wesleyan University, and supported by the 
Keck Northeast Astronomy Consortium student exchange program, made major contributions
to the survey: Michael Santos, Laura Brenneman, Erin Condy, and Kerrie 
McKinstry.  In addition, Wesleyan University students Scott Randall, Nick 
Harrison, Katherine Rhode, Karen Kinemuchi, Kristin Kearns, Eli Beckerman, 
and Janice Lee, plus Cheshire (CT) High School teacher Julie Barker, worked on 
various aspects of the survey.  Without the participation of all these 
dedicated young scientists, this work would not have been as successful. 
We are grateful to Vicki Sarajedini, who assisted in so many ways during 
the final production of this paper.
We thank Bruce Margon and Eric Deutsch for collaborating on the carbon star 
project, including obtaining spectra for several of the candidates.  We also
thank the numerous colleagues with whom we have discussed the KISS project
over the past several years, including Jesus Gallego, Rafael Guzman, David Koo,
and Danial Kunth.  Helpful suggestions made by an anonymous referee are
acknowledged with gratitude.  Finally, we wish to thank the support 
staff of Kitt Peak National Observatory for maintaining the telescope and 
instrument during the early years of the project, and the Astronomy Department 
of Case Western Reserve University for taking over this role after 1997.
In particular, special thanks to Bill Schoening, whose dedication has enabled so 
much useful science to come out of the Burrell Schmidt, and to Heather Morrison
and Paul Harding, who have ensured that the telescope will continue to provide
excellent science for years to come.

The KPNO International Spectroscopic Survey is dedicated to the memory of
Valentin Lipovetsky.




%
%

\clearpage
\begin{deluxetable}{lcccccr}
\tablecaption{Partial List of Previous Schmidt/Objective-Prism Surveys \label{table:elgsurv}}
\tablenum{1}
\tablehead{
\colhead{Survey Name} & \colhead{Type\tablenotemark{a}} & \colhead{Area} & \colhead{Number} & \colhead{Density} & \colhead{Completeness\tablenotemark{b}} & \colhead{Ref.} \\
\colhead{} & \colhead{} & \colhead{(deg.$^2$)} & \colhead{of objects} & \colhead{(\#\ / deg.$^2$)} & \colhead{Limit (m$_B$)}\\
\colhead{(1)} & \colhead{(2)} & \colhead{(3)} & \colhead{(4)} & \colhead{(5)} & \colhead{(6)} & \colhead{(7)}}
\startdata
Haro & C & Lots & $\sim$40 & small & ?? & 1 \\
Kiso & C & 5100 & 8162 & 1.60 & 16.0 & 2 \\
Montreal & C & 4400 & 469 & 0.11 & 14.8 & 3 \\
\\
Markarian & UV & 15000 & 1500 & 0.10 & 15.2 & 4 \\
\\
Tololo & L & 1225 & 201 & 0.16 & -- & 5 \\
UM & L & 667 & 349 & 0.52 & (16.9) & 6 \\
Wasilewski & L & 825 & 96 & 0.18 & (15.2) & 7 \\
POX & L & 82 & 23 & 0.28 & (16.0) & 8 \\
ESO - H$\alpha$ & L & 400 & 113 & 0.28 & -- & 9 \\
UCM & L & 471 & 263 & 0.56 & ($\sim$16.5) & 10 \\
Hamburg & L & 1248 & 196 & 0.16 & -- & 11 \\
Marseille & L & 46.5 & 92 & 1.98 & ($\sim$16.0) & 12 \\
\\
Case & UV + L & 1440 & 1551 & 0.94 & 16.0 & 13 \\
SBS & UV + L & $\sim$990 & $\sim$1300 & 1.31 & ($\sim$17.0) & 14 \\
\\
KISS - red & L & 62.2 & 1128 & 18.14 & (18.1) & 15 \\
KISS - blue & L & 116.6 & 223 & 1.91 & (18.2) & 15 \\
\enddata
\tablenotetext{a}{\thinspace C = Color Selected; UV = UV Excess Selected; L = Line Selected}
\tablenotetext{b}{\thinspace Values in parentheses are median apparent magnitudes for that
survey, not a completeness limit.}
\tablerefs{1 = Haro 1956; 2 = Takase \& Miyauchi-Isobe 1983; 3 = Coziol \etal 1993, 1997; 
4 = Markarian 1967, Markarian, Lipovetskii \& Stepanian 1981; 5 = Smith 1975, Smith, 
Aguirre \& Zemelman 1976; 6 = MacAlpine, Smith \& Lewis 1977, MacAlpine \& Williams 1981; 
7 = Wasilewski 1983; 8 = Kunth, Sargent \& Kowal 1981; 9 = Wamsteker \etal 1985; 
10 = Zamorano \etal 1994, 1996; 11 = Popescu \etal 1996; 12 = Surace \& Comte 1998; 
13 = Pesch \& Sanduleak 1983, Stephenson, Pesch, \& MacConnell 1992; 14 = Markarian 
\etal 1983, Markarian \& Stepanian 1983, Stepanian 1994; 15 = this paper, 
Salzer \etal 2000a,b}
\end{deluxetable}

\end{document}